\documentclass[preprint,12pt]{elsarticle}
\usepackage{amsthm,amsmath,amsfonts,amssymb}
\usepackage{epsfig}
\DeclareMathOperator{\sn}{sn}
\DeclareMathOperator{\cn}{cn}
\DeclareMathOperator{\am}{am}

\journal{International Journal of Mass Spectrometry}
\begin{document}
\begin{frontmatter}
\title{\bf Relativistic shifts of eigenfrequencies\\ in an ideal Penning trap}
\author{Yurij  Yaremko} 
\ead{yar@icmp.lviv.ua}
\address{%
Institute for Condensed Matter Physics of NAS of Ukraine,\\
1 Svientsitskii St., 79011 Lviv, Ukraine
   }%

\begin{abstract}
First-order perturbative calculation of the frequency-shifts caused
by special relativity is performed for a charged particle confined
in a Penning trap. The perturbed motion is approximated by the Jacobian
elliptic functions which describe the periodic orbit repeating itself
sinuously with a period that exceeds $2\pi$. We find relativistic
corrections to amplitudes of oscillating modes as well as shifts of
eigenfrequencies which depend on amplitudes. Besides we find the
relativistic contributions to modular angles. In the low-energy limit
the deformed orbit simplifies to the well-known combination of axial
oscillation and in-plane motion consisting of two circular modes.
We compare the results with the model of relativistic frequency-shifts
developed by J. Ketter {\em et al.}
\cite{Kett14}.
\end{abstract}
\end{frontmatter}
\maketitle

\section{Introduction}\label{Intro}

An ideal Penning trap consists of three electrodes: a ring
electrode and two endcaps \cite[Figs.1,2]{BG86}. Ideally these
electrodes are hyperboloids of revolution, producing a quadrupole
electrostatic potential. A strong homogeneous magnetic field is
oriented strictly along the $z$-axis, i.e. the axis of rotational
symmetry of the electrodes. Even small imperfections of the geometry
of the electrodes and tiny misalignment or inhomogeneity of the
magnetic field yield shifts of particle's eigenfrequencies. Since the
imperfections are experimentally inevitable, they should be made
negligible by means of the most careful design. However, it is
impossible to switch off effects of special relativity. Relativistic
shifts to the energy levels and eigenfrequencies are taken into
account in the modern measurements of the electron magnetic moment
and the fine structure constant \cite{HFG08}. Such shifts are
important in the experiment \cite{Gpp95} where charge-to-mass
ratios for the antiproton and proton are measured with high
precision in order to check CPT invariance. An anharmonic cyclotron
resonance \cite{Kaplan,GDK85} shows that even small nonlinearities
in the electron's motion arising from relativistic corrections
lead to bistable hysteresis.

In Ref.~\cite{Kett14} the authors calculate relativistic corrections
with the help of the perturbation theory developed in Ref.
\cite{Ket14} for anharmonicities of electric and magnetic fields
caused by unavoidable imperfections of the trap's design. A
perturbed trajectory has been parameterized by trigonometric
functions. Relativistic corrections to frequencies of oscillating
modes have been derived.

Recently \cite{YPM15} the dynamics of a charged particle in the
relativistic domain has been studied without any approximation. The
quartic terms appear in effective potential due to special relativity.
We exploit the invariance of the dynamical system with respect to
rotation around the $z$-axis. The symmetry yields the conservation
of the third component of canonical angular momentum. If the conserved
quantity is equal to zero a charge moves along the symmetry axis of
the trap, see \cite[Appendix B]{YPM15}. The axial symmetry allows us
to reduce the dynamics to two degrees of freedom. The oscillating modes,
radial and axial, are joined by the quartic cross term. The term
provides that an energy is continuously being exchanged between these
modes and the system resembles a chaotic double pendulum \cite{BB}.
Indeed, Poincar\'e sections \cite[Figs.~5,6,16]{YPM15} demonstrate
coexistence of regular and chaotic dynamics. The term ``coexistence''
means that the character of particle's orbit strictly depends on the
initial data. Figures 7--10, 12 and 13 in Ref. \cite{YPM15} demonstrate
that the charge follows either chaotic or quasi-periodic trajectory.
Moreover, periodic orbits exist if the initial data take very specific
values which can be revealed from analysis of the Poincar\'e sections.

Under the usual operating conditions of a Penning trap the particle's
velocity $v$ is much less than the speed of light $c$. Typically, the ratio
$v^2/c^2$ is less than $10^{-6}$ \cite{BG86}. In the present paper we suppose
that the total energy is much smaller than the rest energy of the particle.
If the order of magnitude of energy is as in a Penning trap typically,
the cross term is proportional to infinitesimal coefficient and the regular
dynamics dominates. The Poincar\'e section \cite[Fig. 11]{YPM15}
illustrates the situation.

The relativistic quartic terms are similar to those arising from an
octupolar perturbation of the standard electrostatic quadrupole
potential \cite{Ket14,Lara2004}. Poincar\'e sections pictured in
\cite[Fig.~13]{Lara2004} demonstrate that the nonlinear effects caused by
the octupolar perturbation are very similar to those sourced from the
relativistic corrections. In contrast to the tunable imperfection,
the relativistic terms do not constitute a harmonic polynomial
satisfying the Laplace equation. In the present paper we
introduce the electrostatic octupolar potential that cancels
the cross term. As a consequence the energy of oscillating mode,
either axial or radial, is preserved separately. The variables are
separated and the effects of special relativity can be counted
exactly in terms of elliptic integrals and Jacobian elliptic functions
\cite{AbrStg}.

The paper is organized as follows. In Section \ref{non-Relat} we
consider the non-relativistic motion of a charged particle in
an ideal hyperboloid Penning trap. As the dynamical system is
invariant with respect to rotation around the $z$-axis, the
cylindrical coordinates and rotating reference frame are a good
choice \cite{Kret91}. In Section \ref{Relat} we generalize the
results to the relativistic domain. We present all the necessary
information about the relativistic dynamics of a charge in a
Penning trap which is studied in Ref. \cite{YPM15} in details.
In Section \ref{QuasiR} we propose appropriate equations of motion
where axial and radial variables are separated. We find the
relativistic corrections to frequencies and amplitudes as functions
of ratios of energies of radial and axial modes to the rest energy.
We will use the particle's proper time throughout the paper, never
looking at the laboratory time with the exception of relation between
these evolution parameters. It shows how periodic processes look
in the laboratory frame. In Section \ref{Concl} the results
are discussed and a conclusion is drawn.

\section{Non-relativistic orbits}\label{non-Relat}
\setcounter{equation}{0}

Consider the motion of a particle of rest mass $m$ and electric charge $e$ in an ideal Penning trap in non-relativistic approximation. A charge is acted upon the electromagnetic field which is the combination of constant magnetic field and electrostatic field derived from quadrupole potential. A charged particle rotates in a strong homogeneous magnetic field with the so-called cyclotron frequency
\begin{equation}\label{omgC}
\omega_c=\frac{e}{m}B.
\end{equation}
The magnetic field $3$-vector $\mathbf{B}$ is aligned along the positive or negative $z$-axis. For a positive charge $\mathbf{B}=(0,0,B)$, while for a negative one $\mathbf{B}=(0,0,-B)$, so that $\omega_c$ is positive.

The magnetic field confines a charge in the radial $(xy)$-plane only, while the motion along the $z$-axis is unstable. For effective trapping, the magnetic field is superimposed by the electrostatic field produced by three electrodes which are hyperboloids of revolution. Their surfaces are given by the expressions
\begin{equation}\label{elSrf}
z^2=\frac{1}{2}\left(x^2+y^2\right)\pm d^2,
\end{equation}
where $d$ is constant. The upper sign specifies a hyperboloid of two sheets being the surfaces of two end-cap electrodes which have potential $V_0/2$. The lower sign determines a hyperboloid of one sheet encircling the $z$-axis. It is the surface of the ring electrode which has potential $-V_0/2$. Cartesian coordinates $(x,y,z)$ specify the point $\mathbf{x}$ in the rectangular coordinate system with the origin $(0,0,0)$ at the geometric center between electrodes. Defining the axial frequency for a single ion of rest mass $m$ and electric charge $e$
\begin{equation}\label{omgZ}
\omega_z=\sqrt{\frac{e}{m}\frac{V_0}{d^2}},
\end{equation}
the perfect quadrupole electrostatic potential is
\begin{equation}\label{Qpot}
e\Phi(r,z)=\frac{m}{2}\omega_z^2\left[-\frac12\left(x^2+y^2\right)+z^2\right].
\end{equation}

In non-relativistic approximation the motion of a charged particle is governed by the Lagrangian \cite[eq.(4.2)]{Kret91}:
\begin{eqnarray}\label{L0}
L_0&=&\frac{1}{2}m\left(\dot{x}^2+\dot{y}^2+\dot{z}^2\right)-
\frac14m\omega_z^2\left(2z^2-x^2-y^2\right)\nonumber\\
&-&\frac12m\omega_c\left(\dot{x}y-\dot{y}x\right).
\end{eqnarray}
The standard procedure leads to the Hamiltonian
\begin{eqnarray}\label{H0}
H_0&=&\frac{1}{2m}\left(p_x^2+p_y^2+p_z^2\right)
+\frac12\omega_c\left(p_xy-p_yx\right)\nonumber\\
&+&\frac12m\Omega^2\left(x^2+y^2\right)+\frac12m\omega_z^2z^2,
\end{eqnarray}
where frequency $\Omega=\frac12\omega_c\sqrt{1-\kappa}$. Symbol $\kappa$ denotes the ratio
\begin{equation}\label{kapp}
\kappa=\frac{2\omega_z^2}{\omega_c^2},
\end{equation}
which is called a trapping parameter. A charge is confined if the inequality $\kappa<1$ is fulfilled.

As the Lagrangian (\ref{L0}) is invariant with respect to rotation
around the $z$-axis the cylindrical coordinates $(\rho,\varphi,z)$ are
a good choice \cite{Kret91,Kret92}. In these coordinates the
Hamiltonian (\ref{H0}) takes the form
\begin{eqnarray}\label{H0cyl}
H_0&=&\frac{1}{2m}\left(p_\rho^2+\frac{p_\varphi^2}{\rho^2}+p_z^2\right)-\frac12\omega_cp_\varphi\nonumber\\
&+&\frac12m\Omega^2\rho^2+\frac12m\omega_z^2z^2.
\end{eqnarray}
As the polar angle $\varphi$ is cyclic coordinate, the conjugate
momentum $p_\varphi$ is the first integral. The momentum
$p_\varphi$ is the third component $L_z=xp_y-yp_x$ of the canonical
angular momentum $\textbf{L}=\textbf{x}\times\textbf{p}$.

In Ref.~\cite{Kret91} the rotating coordinates are introduced. In
this paper we use privileged reference frame rotating around the
$z$-axis with frequency $\omega_c/2$ in clockwise direction.
The transformation of Cartesian coordinates
\begin{equation}\label{xy_nr}
\left[\begin{array}{c}
x\\[1.2em]
y
\end{array}
\right]= \left[
\begin{array}[c]{cc}
\cos\left(\frac{\displaystyle\omega_c}{\displaystyle 2}t\right)&\sin\left(\frac{\displaystyle\omega_c}{\displaystyle 2}t\right)\\
-\sin\left(\frac{\displaystyle\omega_c}{\displaystyle 2}t\right)&\cos\left(\frac{\displaystyle\omega_c}{\displaystyle 2}t\right)
\end{array}
\right] \left[\begin{array}{c}
\tilde{x}\\[1.2em]
\tilde{y}
\end{array}\right]
\end{equation}
is equivalent to the following time-dependent canonical transformation:
\begin{align}\label{can-rot}
\rho&=\tilde{\rho}\,,& p_\rho&=p_{\tilde{\rho}}\,,\nonumber\\
\varphi&=\tilde{\varphi}-\frac12\omega_ct\,,&
p_\varphi&=p_{\tilde{\varphi}}\,.
\end{align}
Tilted rectangular coordinates are related to the tilted cylindrical coordinates as usual: $\tilde{x}=\tilde{\rho}\cos\tilde{\varphi}$ and $\tilde{y}=\tilde{\rho}\sin\tilde{\varphi}$.

In new coordinates the Hamiltonian does not contain the term which
is proportional to the constant momentum $p_\varphi$\,:
\begin{eqnarray}\label{H-rot}
\tilde{H}_0&=&H_0+\frac12\omega_cp_\varphi\\
&=&\frac{1}{2m}\left(p_{\tilde{\rho}}^2+\frac{p_{\tilde{\varphi}}^2}{\tilde{\rho}^2}\right)+\frac12m\Omega^2\tilde{\rho}^2
+\frac{1}{2m}p_z^2+\frac12m\omega_z^2z^2.\nonumber
\end{eqnarray}
The Hamiltonian (\ref{H-rot}) is the sum of the terms governing the
motion in the $(\tilde{x}\tilde{y})$-plane and the terms defining the motion along
the $z$-axis: $\tilde{H}_0=\tilde{H}_r+\tilde{H}_z$. We denote
corresponding energies as $\tilde{E}_r^{(0)}$ and
$\tilde{E}_z^{(0)}$, respectively. In axial direction, the charge oscillates
around zero equilibrium point with frequency $\omega_z$.
Besides the axial frequency, the axial orbit
\begin{equation}\label{axial-Nr}
z(t)=A_z\sin\left(\omega_zt-\phi_z\right)
\end{equation}
contains two constants: one, the scaled axial energy $\tilde{{\cal
E}}_z^{(0)}=\tilde{E}_z^{(0)}/m$, defines the amplitude $A_z=\sqrt{2\tilde{{\cal
E}}_z^{(0)}}/\omega_z$, the other, $\phi_z$, appears as a phase shift.

\begin{figure}[ht]
\begin{center}
\includegraphics*[scale=0.7,angle=0,trim=5 5 5 5]{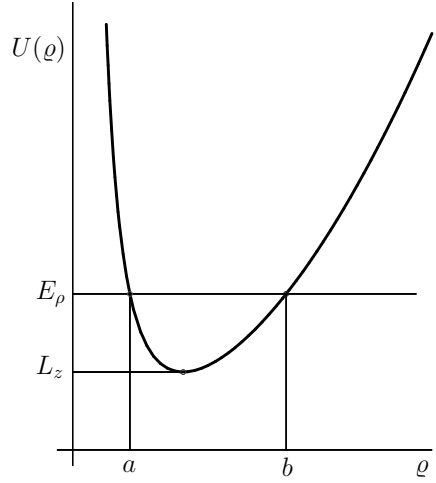}
\end{center}
\caption{Graph of the effective potential (\ref{Unr}) for a fixed
value $L_z$ of the momentum $p_{\tilde{\varphi}}$. For a fixed energy
$E_\rho=\Omega^{-1}\tilde{E}_r^{(0)}$
the radial variable $\varrho$ oscillates between points $a$ and
$b$. At turning points the radial velocity $\dot{\varrho}=0$. At
the point of minimum $\varrho_{\rm min}=\sqrt{L_z}$ the normalized energy
$E_\rho=L_z$.\label{U-nr}}
\end{figure}

To simplify the radial Hamiltonian $\tilde{H}_r$ we introduce new canonical variables
\begin{equation}\label{Dless-nr}
\varrho=\sqrt{m\Omega}\tilde{\rho},\qquad
p_\varrho=\frac{p_{\tilde{\rho}}}{\sqrt{m\Omega}},
\end{equation}
and define the dimensionless time $t'=\Omega t$. In these variables the radial Hamiltonian $H_\rho=\Omega^{-1}\tilde{H}_r$ takes the form
\begin{equation}\label{Hr-nr}
H_\rho=\frac12p_\varrho^2+U(\varrho),
\end{equation}
where effective potential $U(\varrho)$ is the sum of centrifugal barrier and potential of harmonic oscillator:
\begin{equation}\label{Unr}
U(\varrho)=\frac12\left(\frac{L_z^2}{\varrho^2}+\varrho^2\right).
\end{equation}
We denote $L_z=p_{\tilde{\varphi}}$. The function is pictured in Fig.~\ref{U-nr}.

The Hamiltonian (\ref{Hr-nr}) itself is the first integral. We
denote $E_\rho\geq L_z$ a fixed energy level. Putting $p_\varrho=\dot{\varrho}$
and factoring the right hand side of the equation
$\dot{\varrho}^2=2\left(E_\rho-U(\varrho)\right)$ we obtain
\begin{equation}\label{radEqnr}
\dot{\varrho}^2=\frac{1}{\varrho^2}\left(b^2-\varrho^2\right)\left(\varrho^2-a^2\right),
\end{equation}
where
\begin{subequations}\label{abA}
\begin{align}
a&=\sqrt{E_\rho\left(1-A\right)},\quad b=\sqrt{E_\rho\left(1+A\right)};\label{abA:ab}\\
A&=\sqrt{1-\frac{L_z^2}{E_\rho^2}}.\label{abA:A}
\end{align}
\end{subequations}
The parameter $0\leq A<1$. The solution of eq. (\ref{radEqnr}) is
\begin{equation}\label{r2nr}
\varrho^2(t')=a^2+\left(b^2-a^2\right)\sin^2\left(t'-\phi_r\right),
\end{equation}
where $\phi_r$ is an initial phase. The inequality $E_\rho\geq L_z$
places stringent requirement on the axial energy too. Indeed, the
energy in laboratory frame is less than the energy in the rotating
frame: $E_r^{(0)}=\tilde{E}_r^{(0)}-\omega_cL_z/2$ (see
eq.~(\ref{H-rot})). The axial energy $E_z^{(0)}=\tilde{E}_z^{(0)}$
should at least compensate the negative minimal energy $E_{r,{\rm
min}}^{(0)}=\Omega
L_z-\omega_cL_z/2=-\frac12\omega_c\left(1-\sqrt{1-\kappa}\right)L_z$.
Therefore,
\begin{equation}\label{Ez-min}
E_z^{(0)}\geq\frac12\omega_c\left(1-\sqrt{1-\kappa}\right)L_z.
\end{equation}

The angular velocity can be obtained once the radial orbit is known:
\begin{equation}\label{polar-Nr}
\frac{{\rm d}\tilde{\varphi}}{{\rm
d}t}=\frac{1}{m}\frac{p_{\tilde{\varphi}}}{\tilde{\rho}^2}.
\end{equation}
In terms of new variables (\ref{Dless-nr}) the equation
takes the form
\begin{equation}\label{polar-angle}
\frac{{\rm d}\tilde{\varphi}}{{\rm d}t'}=\frac{L_z}{\varrho^2}.
\end{equation}
Inserting the solution (\ref{r2nr}) we derive the polar angle
\begin{equation}\label{varphi-nr}
\tilde{\varphi}(t')=\varphi_0+\arctan\left[\frac{b}{a}\tan\left(t'-\phi_r\right)\right],
\end{equation}
after integration over the evolution parameter $t'$.

\begin{figure}[ht]
\begin{center}
\includegraphics*[scale=0.7,angle=0,trim=2 2 0 1]{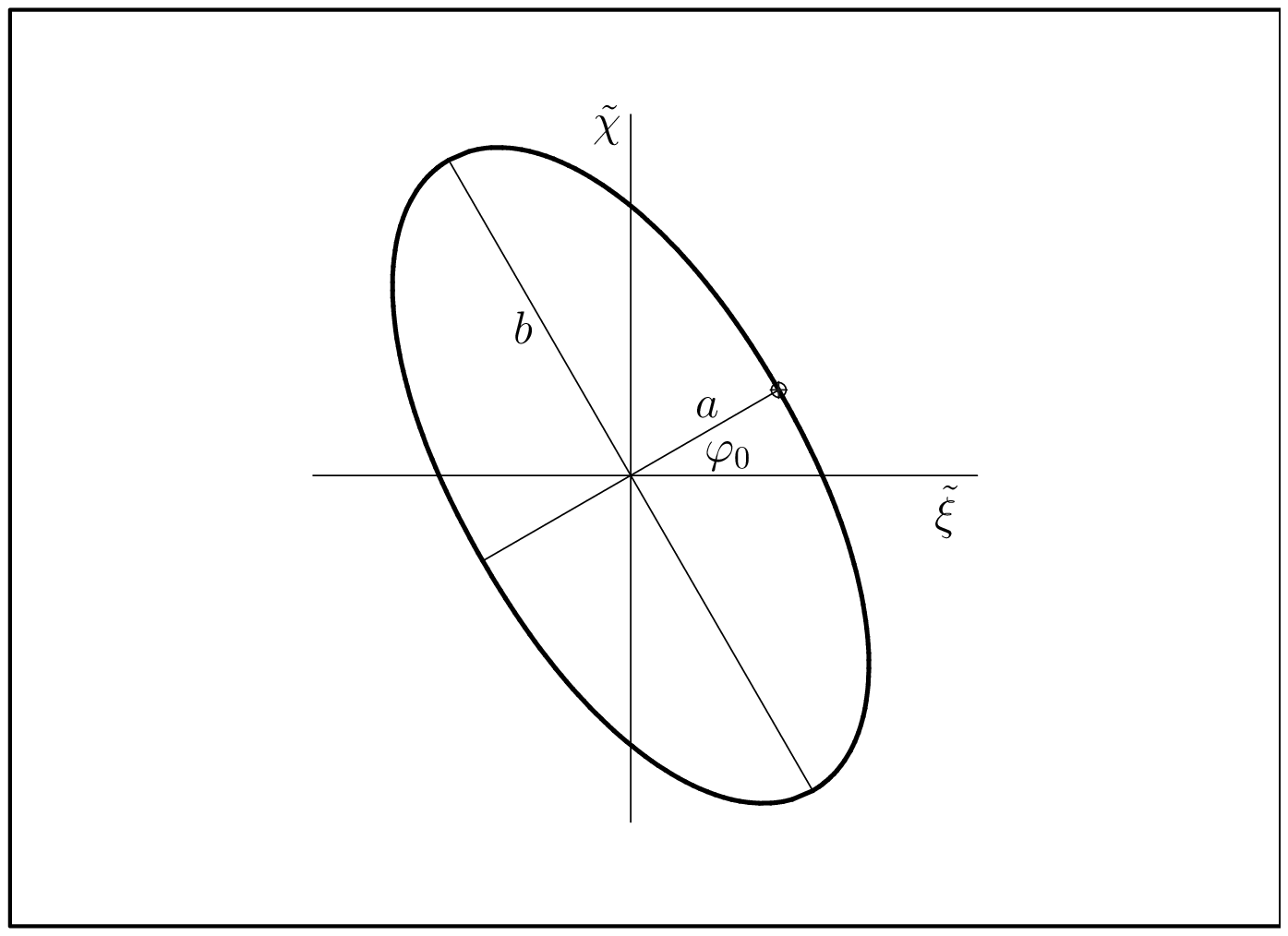}
\end{center}
\caption{In the rotating reference frame a charge moves along the
ellipse with minor semi-axis $a$ and major semi-axis $b$ given in
eqs.~(\ref{abA:ab}). The tilt angle $\varphi_0$ arises in
eq.~(\ref{varphi-nr}) as constant of integration of angular velocity.
A charge follows the same perfect ellipse constantly: the starting
point (circle) coincides with the point after period of
oscillation (cross).
\label{Ellipse}}
\end{figure}

To visualize the orbit in the plane which is orthogonal to the $z$-axis we come back to rectangular
coordinates $\tilde{\xi}(t')=\varrho(t')\cos\tilde{\varphi}(t')$ and
$\tilde{\chi}(t')=\varrho(t')\sin\tilde{\varphi}(t')$. Using the
identities
\begin{equation}\label{cssn-arctn}
\cos\arctan h=\frac{1}{\sqrt{1+h^2}},\quad \sin\arctan
h=\frac{h}{\sqrt{1+h^2}},
\end{equation}
after some algebra and trigonometry, we obtain
\begin{equation}\label{xy-nr}
\left[\begin{array}{c}
\tilde{\xi}(t')\\
\tilde{\chi}(t')
\end{array}
\right]= \left[
\begin{array}[c]{cc}
\cos\varphi_0&-\sin\varphi_0\\
\sin\varphi_0&\cos\varphi_0
\end{array}
\right] \left[\begin{array}{c}
a\cos\left(t'-\phi_r\right)\\
b\sin\left(t'-\phi_r\right)
\end{array}\right].
\end{equation}
In the rotating frame (\ref{xy_nr}) the in-plane orbit is an ellipse
with center at the origin $(0,0)$, minor semi-axis $a$ and major semi-axis
$b$ (see Fig.~\ref{Ellipse}). The eccentricity of the ellipse is
$$
\varepsilon=\sqrt{\frac{2A}{1+A}}.
$$
where $A$ is given in eq.~(\ref{abA:A}). At the minimum of the potential (\ref{Unr}) the energy $E_\rho$ is equal to the third component of the canonical angular momentum $L_z$ and the eccentricity is equal to zero. The charge follows the circle with radius $\sqrt{L_z}$.

\begin{figure}[ht]
\begin{center}
\includegraphics*[scale=0.7,angle=0,trim=2 2 0 1]{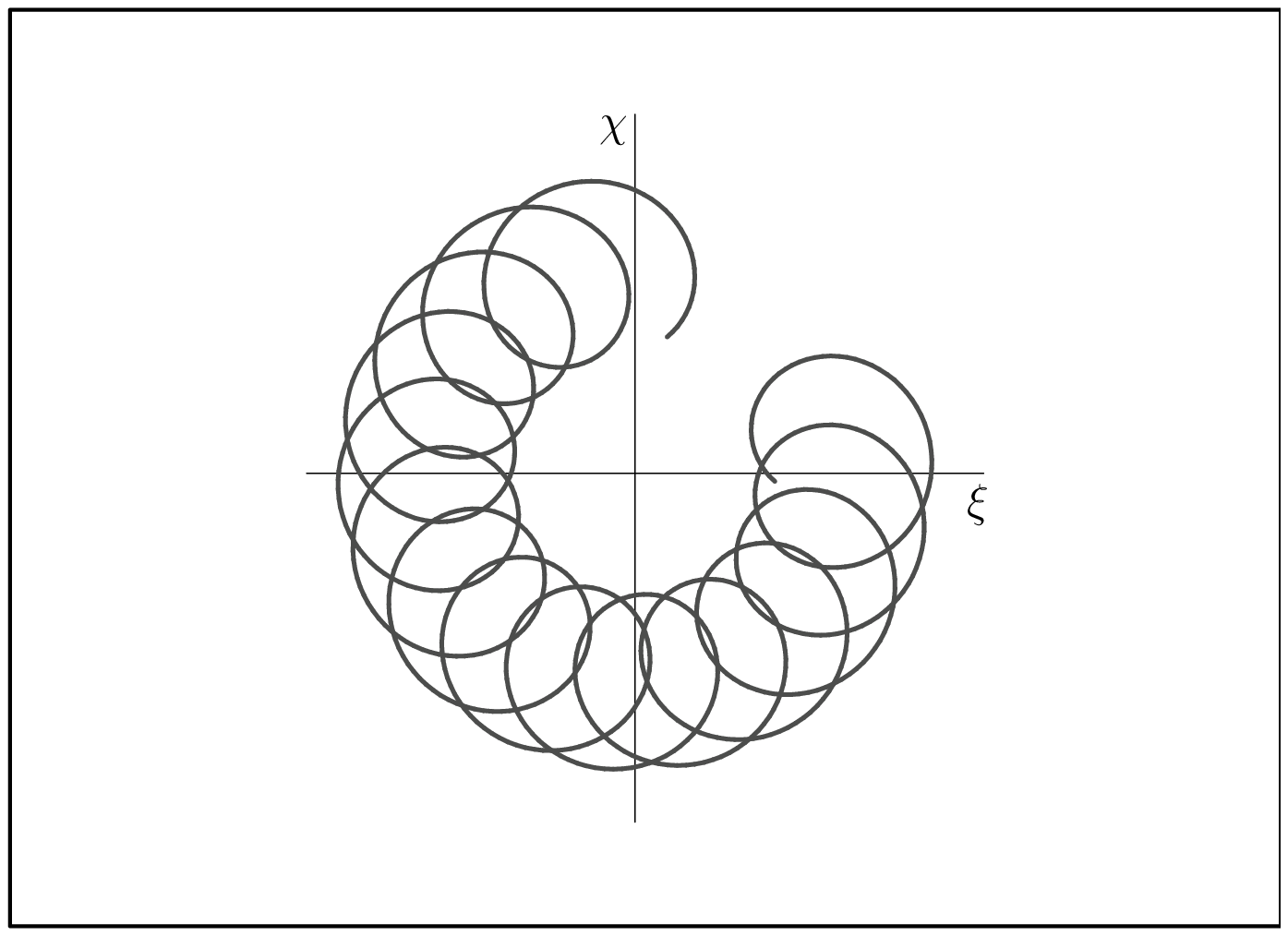}
\end{center}
\caption{In the inertial reference frame a charge moves on epicyclic orbit which consists of a fast circular
motion with small radius $(b-a)/2$ carried along by a slow circular motion with radius $(b+a)/2$.The orbit was known in ancient Greece as {\it epitrochoid}.
\label{Rosette}}
\end{figure}

What is the form of this trajectory in the laboratory frame of reference? In terms of dimensionless time $t'=\Omega t$ and variables $\tilde{\xi}$ and $\tilde{\chi}$ the transformation (\ref{xy_nr}) looks as
\begin{equation}\label{xy-inrt}
\left[\begin{array}{c}
\xi(t')\\[1.5em]
\chi(t')
\end{array}
\right]= \left[
\begin{array}[c]{cc}
\cos\frac{\displaystyle t'}{\displaystyle\sqrt{1-\kappa}}&\sin\frac{\displaystyle t'}{\displaystyle\sqrt{1-\kappa}}\\
-\sin\frac{\displaystyle
t'}{\displaystyle\sqrt{1-\kappa}}&\cos\frac{\displaystyle
t'}{\displaystyle\sqrt{1-\kappa}}
\end{array}
\right] \left[\begin{array}{c}
\tilde{\xi}(t')\\[1.5em]
\tilde{\chi}(t')
\end{array}\right].
\end{equation}
Inserting the solutions (\ref{xy-nr}) we obtain the following combination
of two oscillating modes:
\begin{eqnarray}
\xi(t')&=&-\frac12\left(b-a\right)\cos\left[\left(\frac{\displaystyle
1}{\displaystyle\sqrt{1-
\kappa}}+1\right)t'-\phi_+\right]\nonumber\\
&+&\frac12\left(b+a\right)\cos\left[\left(\frac{\displaystyle
1}{\displaystyle\sqrt{1-
\kappa}}-1\right)t'-\phi_-\right],\label{x-tau}\\
\chi(t')&=&\frac12\left(b-a\right)\sin\left[\left(\frac{\displaystyle
1}{\displaystyle\sqrt{1-
\kappa}}+1\right)t'-\phi_+\right]\nonumber\\
&-&\frac12\left(b+a\right)\sin\left[\left(\frac{\displaystyle
1}{\displaystyle\sqrt{1- \kappa}}-1\right)t'-\phi_-\right].\label{y-tau}
\end{eqnarray}
The phases $\phi_+=\varphi_0+\phi_r$ and $\phi_-=\varphi_0-\phi_r$
are composed from phase shifts arising in the solution (\ref{r2nr})
of radial equation of motion and in the polar angle orbit
(\ref{varphi-nr}). The particle moves within a circular strip with
outer radius $b$ and inner radius $a$ which are given in eqs.~(\ref{abA:ab}).
The rosette shaped curve is pictured in Fig.~\ref{Rosette}.

For our subsequent considerations it is again of interest to restore
the dimension variables
$$
\left.x(t)=\frac{\xi(t')}{\sqrt{m\Omega}}\right|_{t'=\Omega t},\quad
\left.y(t)=\frac{\chi(t')}{\sqrt{m\Omega}}\right|_{t'=\Omega t}.
$$
It is convenient to introduce new parameters: scaled energy
$\tilde{{\cal E}}_r^{(0)}=\tilde{E}_r^{(0)}/m$ and scaled angular
momentum $\ell^{(0)}=L_z/m$. The in-plane orbit is a rosette shaped
curve around the center $(0,0)$:
\begin{align}
x(t)&=-R_+\cos\left(\omega_+t-\phi_+\right)+R_-\cos\left(\omega_-t-\phi_-\right),\nonumber\\
y(t)&=R_+\sin\left(\omega_+t-\phi_+\right)-R_-\sin\left(\omega_-t-\phi_-\right).\label{epitrochoid}
\end{align}
Here $\omega_+=\frac12\omega_c\left(1+\sqrt{1-\kappa}\right)$ is the reduced cyclotron frequency and
$\omega_-=\frac12\omega_c\left(1-\sqrt{1-\kappa}\right)$ is the magnetron frequency.

A charge moves on epicyclic orbit which consists of a fast circular
cyclotron motion with a small radius
$$
R_+=\frac{\sqrt{\tilde{{\cal
E}}_r^{(0)}}}{2\Omega}\left(\sqrt{1+A}-\sqrt{1-A}\right),
$$
carried along by a slow circular magnetron motion with a large radius
$$
R_-=\frac{\sqrt{\tilde{{\cal
E}}_r^{(0)}}}{2\Omega}\left(\sqrt{1+A}+\sqrt{1-A}\right).
$$
In this parametrization the constant (\ref{abA:A}) looks as follows
\begin{equation}\label{A0}
A=\sqrt{1-\left(\frac{\ell^{(0)}\Omega}{\tilde{{\cal E}}_r^{(0)}}\right)^2}.
\end{equation}
We will compare orbits perturbed by relativistic effects with those in this simple model.

\section{Relativistic dynamics}\label{Relat}
\setcounter{equation}{0}

In this Section we draw a rough sketch of results presented in
Ref.~\cite{YPM15} where dynamics of a single ion in the Penning trap
in the relativistic framework without approximations is analyzed.
We suppose that the charged particle moves along the time-like world line
parameterized by four functions either rectangular Cartesian coordinates
$(x^0(\tau),x^k(\tau))$, or cylindrical coordinates
$(x^0(\tau),\rho(\tau),\varphi(\tau),z(\tau))$, of the proper time
$\tau$. The dynamics in relativistic domain is governed by the
Lorentz force equation $m\dot{u}^\alpha=eF^\alpha{}_\beta u^\beta$ where
$u^\beta={\rm d}x^\beta(\tau)/{\rm d}\tau$ is particle's four-velocity and
$\dot{u}^\alpha={\rm d}u^\alpha(\tau)/{\rm d}\tau$ is its four-acceleration.
The electromagnetic field tensor $\hat{F}$ \cite[Eq.(12)]{YPM15} is the combination of
constant magnetic field and electric field derived from quadrupole
potential. To put the Lorentz force equation
into Lagrangian framework \cite[Eq.(15)]{YPM15} we parameterize
the world line by an arbitrary evolution parameter $\lambda$.
After that we transform the Lagrangian using cylindrical coordinates
$x^1=\rho\cos\varphi$, $x^2=\rho\sin\varphi$, $x^3=z$ relative to geometrical
center between electrodes:
\begin{equation}\label{Lagr-cyl}
L=-m\gamma^{-1}-e\Phi(\rho,z)\dot{x}^0+\frac{m}{2}\omega_c\rho^2\dot{\varphi}.
\end{equation}
In this Lagrangian the inverse Lorentz factor
\begin{equation}\label{gamm}
\gamma^{-1}=\sqrt{\left(\frac{{\rm d}x^0}{{\rm d}\lambda}\right)^2-\left(\frac{{\rm d}\rho}{{\rm d}\lambda}\right)^2-\rho^2\left(\frac{{\rm d}\varphi}{{\rm d}\lambda}\right)^2-\left(\frac{{\rm d}z}{{\rm d}\lambda}\right)^2},
\end{equation}
and the quadrupole potential (\ref{Qpot}) are expressed in terms of cylindrical coordinates.

Variation of the action integral $S=\int{\rm d}\lambda L$ yields
equations of motion. There are two first integrals which correspond
to two cyclic coordinates, $x^0$ and $\varphi$:
\begin{subequations}\label{First}
\begin{align}
p_0&=-m\gamma\frac{{\rm d}x^0}{{\rm d}\lambda}-e\Phi(\rho,z),\label{First:E}\\
p_\varphi&=m\rho^2\gamma\frac{{\rm d}\varphi}{{\rm d}\lambda}+\frac{m}{2}\omega_c\rho^2.\label{First:L}
\end{align}
\end{subequations}
Obviously, $p_0$ is the sum of kinetic energy and
potential energy taken with opposite sign, i.e., $p_0=-E$.
The momentum $p_\varphi$ canonically conjugated to the polar
angle $\varphi$ is the third component of the relativistic angular momentum
$\textbf{L}=\textbf{x}\times\textbf{p}$.

To simplify the expressions we restore the proper time
parametrization ${\rm d}\tau=\gamma^{-1}{\rm d}\lambda$. From the
conserved quantities (\ref{First}) one can easily derive the
relations
\begin{subequations}\label{u0vphi}
\begin{align}
\dot{x}^0&={\cal E}-\frac{1}{2}\omega_z^2\left(-\frac12\rho^2+z^2\right),\label{u0vphi:E}\\
\dot{\varphi}&=\frac{\ell}{\rho^2}-\frac{\omega_c}{2},\label{u0vphi:L}
\end{align}
\end{subequations}
where ${\cal E}=E/m$ and $\ell=p_\varphi/m$ and the overdot means
differentiation with respect $\tau$.

If we choose $\tau$ as the evolution parameter the equations of
motion of the radial and axial variables take the form
\begin{equation}\label{startEqs}
  \ddot{\rho}=\rho\left(\frac{\omega_z^2}{2}u^0+\dot{\varphi}^2+\omega_c\dot{\varphi}\right),\quad
  \ddot{z}=-\omega_z^2zu^0,
\end{equation}
where $u^0\equiv\dot{x}^0$ is the zeroth component of particle's 4-velocity. In
this parametrization the norm of particle's four-velocity is equal
to $-1$:
$$
-(\dot{x}^0)^2+\dot{\rho}^2+\rho^2\dot{\varphi}^2+\dot{z}^2=-1.
$$
Substituting the right-hand side of eq.~(\ref{u0vphi:E}) for $\dot{x}^0$
we obtain
\begin{equation}\label{unitN}
\dot{\rho}^2+\rho^2\dot{\varphi}^2+\dot{z}^2+2{\cal
E}\left(\frac{e}{m}\Phi\right)-\left(\frac{e}{m}\Phi\right)^2={\cal
E}^2-1.
\end{equation}
Inserting eq.~(\ref{u0vphi:L}) we derive that in reference to the
privileged rotating frame (\ref{can-rot}) the unit norm velocity
condition takes the form
\begin{eqnarray}
\left(\frac{{\rm d}\tilde{\rho}}{{\rm d}\tau}\right)^2&+&\frac{\ell^2}{\tilde{\rho}^2}+\frac{\omega_c^2}{4}\tilde{\rho}^2
+\left(\frac{{\rm d}z}{{\rm d}\tau}\right)^2+2{\cal
E}\left(\frac{e}{m}\Phi\right)-\left(\frac{e}{m}\Phi\right)^2\nonumber\\
&=&{\cal E}^2-1+\ell\omega_c.\label{unitUrot}
\end{eqnarray}
In analogy with the right-hand side of identity (\ref{unitN}) we denote the
energy level in the rotating reference frame as $(\tilde{{\cal E}}^2-1)/2$, so that
\begin{equation}\label{EE}
\tilde{{\cal E}}^2-1={\cal E}^2-1+\ell\omega_c.
\end{equation}
According to eqs. (\ref{can-rot}), the radial coordinate and radial canonical momentum in the rotating reference frame coincide with their counterparts in the laboratory reference frame. For this reason we do not mark these coordinates by the sign ``tilde'' further in this Section.

Substituting the right-hand sides of eqs.~(\ref{u0vphi}) for $u^0$ and
$\dot{\varphi}$ in eqs. (\ref{startEqs}) we derive the equations of
motion describing the dynamical system with two degrees of freedom:
\begin{subequations}\label{rhoz}
\begin{align}
\ddot{\rho}&=-\Omega_\epsilon^2\rho+\frac{\ell^2}{\rho^3}+\frac{\omega_z^4}{8}\rho^3-\frac{\omega_z^4}{4}\rho z^2,\label{rhoz:r}\\
\ddot{z}&=-{\cal E}\omega_z^2z+\frac{\omega_z^4}{2}z^3-\frac{\omega_z^4}{4}z\rho^2,\label{rhoz:z}
\end{align}
\end{subequations}
where the relativistic radial frequency $\Omega_\epsilon=\frac12\omega_c\sqrt{1-{\cal E}\kappa}$. The shift of this frequency is caused by the relativistic mass increase \cite{BG86,GDK85}. The system of the second order differential equations can be put
into Hamiltonian framework
\begin{equation}\label{Ham}
\tilde{H}=\frac12p_\rho^2+\frac12p_z^2+V(\rho,z),
\end{equation}
with potential
\begin{equation}\label{V}
V(\rho,z)=\frac12\left[\Omega_\epsilon^2\rho^2 +
\frac{\displaystyle\ell^2}{\displaystyle \rho^2}+{\cal
E}\omega_z^2z^2-\frac{\displaystyle\omega_z^4}{\displaystyle
4}\left(\frac14\rho^4-\rho^2z^2+z^4\right)\right].
\end{equation}
The potential consists of the modified quadrupole potential,
centrifugal barrier, and quartic terms originating from the special relativity.
The oscillating modes are coupled by the cross term $\omega_z^4\rho^2z^2/8$. The
Hamiltonian (\ref{Ham}) is also the conserved quantity. As the left-hand side of the
velocity norm condition (\ref{unitUrot}) can be expressed as the double Hamiltonian
(\ref{Ham}), the energy level is equal to $\left(\tilde{{\cal E}}^2-1\right)/2$.

The Hamiltonian (\ref{Ham}) produces two second-order differential
equations (\ref{rhoz}) on variables $\rho$ and $z$. Once the radial orbit
$\rho(\tau)$ is known, one can find out $\varphi(\tau)$
integrating the first integral (\ref{u0vphi:L}). Substituting
the orbits $\rho(\tau)$ and $z(\tau)$ in the integral of motion
(\ref{u0vphi:E}) and integrating the first order differential equation we derive the laboratory time
$x^0(\tau)$ as function of the proper time $\tau$. The equations (\ref{rhoz})
describe two oscillating modes which are related to each other.
In the next Section we separate variables in the quasi-relativistic
approximation of these equations by means of precisely tuned octupolar
perturbation of the perfect quadrupole potential.

\section{Quasi-relativistic approximation}\label{QuasiR}
\setcounter{equation}{0}

In this Section we find the small relativistic corrections to
non-relativistic orbits. We are interested in the orbits of low
energetic particles for which relativistic effects play an important
role. We restore the speed of light $c$ in Hamiltonian (\ref{Ham}).
We replace the frequencies $\omega_c$ and $\omega_z$ by $\omega_c/c$
and $\omega_z/c$, respectively. We substitute $p_\rho/mc$ for
$p_\rho$ and $p_z/mc$ for $p_z$. The scaled angular momentum
$\ell$ is also replaced by $\ell/c$ where the numerator is
\begin{equation}\label{elll}
\ell=\ell^{(0)}+\frac{1}{c^2}\ell^{(1)}.
\end{equation}
The first term, $\ell^{(0)}$, is just the non-relativistic constant
of motion involved in eq.~(\ref{A0}). As the Hamiltonian (\ref{Ham})
governs the dynamics in the rotating reference frame where energy level
is $\left(\tilde{{\cal E}}^2-1\right)/2$ we restore the sign ``tilde''
over the radial coordinates. With the precision sufficient
for our purposes we write the scaled energy $\tilde{{\cal E}}=\tilde{E}/mc^2$ as
\begin{equation}\label{calEtld}
\tilde{{\cal E}}=1+\frac{1}{c^2}\tilde{{\cal E}}^{(0)}+\frac{1}{c^4}\tilde{{\cal E}}^{(1)}.
\end{equation}
As the level of energy we obtain the expression
\begin{equation}\label{calE-1}
\frac12\left(\tilde{{\cal E}}^2-1\right)=
\frac{1}{c^2}\left(\tilde{{\cal E}}^{(0)}+\frac{1}{c^2}\tilde{{\cal E}}^{(1)}\right)
\left(1+\frac{1}{2c^2}\tilde{{\cal E}}^{(0)}\right),
\end{equation}
which prompts that we should overmultiply the equality $\tilde{H}=\left(\tilde{{\cal E}}^2-1\right)/2$
on $mc^2$. After some algebra we obtain the quasi-relativistic Hamiltonian
\begin{eqnarray}\label{H-qrel}
\tilde{H}&=&\frac{1}{2m}\left(p_{\tilde{\rho}}^2+p_z^2\right)+
\frac{m}{2}\left(\Omega_\epsilon^2\tilde{\rho}^2+\frac{\ell^2}{\tilde{\rho}^2}+{\cal E}\omega_z^2z^2\right)\nonumber\\
&-&\frac{m\omega_z^4}{8c^2}\left(\frac14\tilde{\rho}^4-\tilde{\rho}^2z^2+z^4\right),
\end{eqnarray}
with the energy level
\begin{equation}\label{En-Lvl}
\tilde{H}=\tilde{E}^{(0)}\left(1+\frac{1}{2mc^2}\tilde{E}^{(0)}\right)+\frac{1}{c^2}\tilde{E}^{(1)}.
\end{equation}
The squared frequency $\Omega_\epsilon^2=\frac14\omega_c^2(1-{\cal E}\kappa)$ depends on the energy in the laboratory reference frame ${\cal E}=E/mc^2$ which we also express as series in powers $1/c^2$
\begin{equation}\label{calE}
{\cal E}=1+\frac{1}{c^2}{\cal E}^{(0)}+\frac{1}{c^4}{\cal E}^{(1)}.
\end{equation}
From eq. (\ref{EE}) which relates the energies one can easily derive
\begin{eqnarray}
\tilde{{\cal E}}^{(0)}&=&{\cal
E}^{(0)}+\frac12\omega_c\ell^{(0)},\label{EE-EE}\\
\tilde{{\cal E}}^{(1)}&=&{\cal
E}^{(1)}+\frac12\omega_c\ell^{(1)}-\frac12\omega_c\ell^{(0)}\left({\cal
E}^{(0)}+\frac14\omega_c\ell^{(0)}\right).\nonumber
\end{eqnarray}
The frequency $\Omega_\epsilon$ can also be developed in series up to the first order in powers $1/c^2$:
\begin{equation}\label{Omg:eps}
\Omega_\epsilon=\Omega\left(1-\frac{1}{2c^2}L_\kappa{\cal E}^{(0)}\right),\quad L_\kappa=\frac{\kappa}{1-\kappa}.
\end{equation}

We suppose that the design of electrodes is changed intentionally in such a way that they produce the octupolar perturbation to the perfect quadrupole potential. New electrostatic potential is \cite[\S 3.1.]{Ket14}:
\begin{eqnarray}\label{Phi24}
\Phi(r,\theta)&=&\Phi_2(\rho,z)+\Phi_4(\rho,z)\\
&=&C_2\frac{eV_0}{2d^2}\left(z^2-\frac12\rho^2\right)+C_4\frac{eV_0}{2d^4}\left(z^4-3z^2\rho^2+\frac38\rho^4\right).\nonumber
\end{eqnarray}
Putting $C_2=1$ we consider that $\Phi_2(\rho,z)$ is the quadrupole potential (\ref{Qpot}) in terms of cylindrical coordinates. We assume that the dimensionless prefactor $C_4$ takes the value
\begin{equation}\label{C4}
C_4=\frac{1}{c^2}\frac{d^2\omega_z^2}{12}.
\end{equation}
The octupolar potential cancels the cross term in the quasi-relativistic Hamiltonian (\ref{H-qrel}) which generalizes the non-relativistic Hamiltonian (\ref{H-rot}) with perfect quadrupole potential. The dynamics is governed by the Hamiltonian
\begin{eqnarray}\label{H_qrel}
\tilde{H}&=&\frac{1}{2m}\left(p_{\tilde{\rho}}^2+p_z^2\right)+
\frac{m}{2}\left(\Omega_\epsilon^2\tilde{\rho}^2+\frac{\ell^2}{\tilde{\rho}^2}+{\cal E}\omega_z^2z^2\right)\nonumber\\
&-&\frac{m\omega_z^4}{8c^2}\left(\frac18\tilde{\rho}^4+\frac23z^4\right),
\end{eqnarray}
which is the sum of the terms defining the evolution of radial variable and the terms defining the motion along the $z$-axis: $\tilde{H}=H_\rho+H_z$. The radial energy, $\tilde{E}_r$, and axial energy, $\tilde{E}_z$, are the first integrals. Our next task is to solve corresponding equations of motion.

\subsection{Radial motion}\label{Radial}

\begin{figure}[ht]
\begin{center}
\includegraphics*[scale=0.7,angle=0,trim=1 3 2 1]{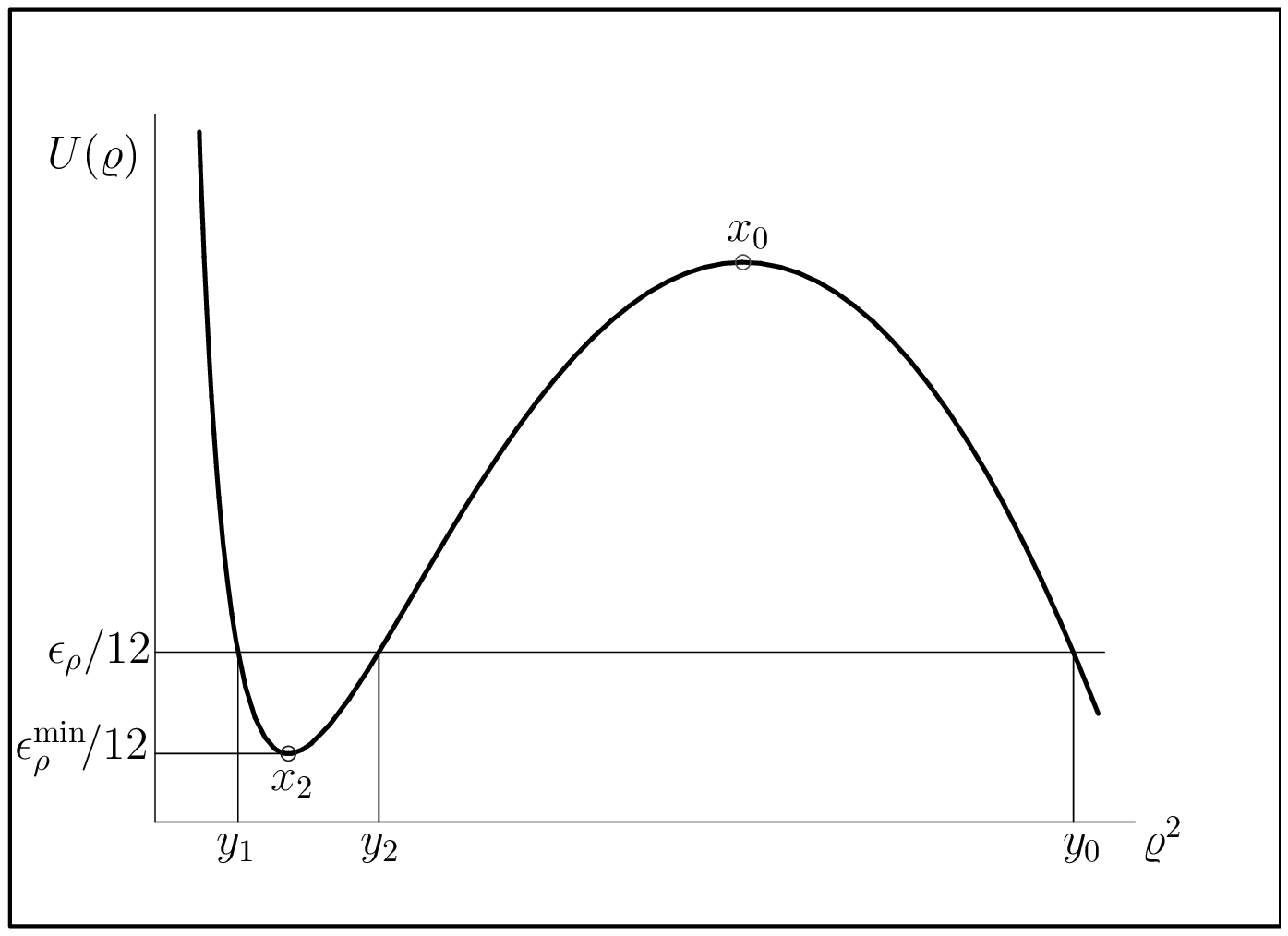}
\end{center}
\caption{Graph of the relativistic potential (\ref{U-rho}) for a
fixed value $\lambda$. For a fixed value $\epsilon_\rho$ the squared
radial variable oscillates between the roots $y_1$ and $y_2$ of the
algebraic equation $U(\varrho)=\epsilon_\rho/12$. The third root of this
cubic polynomial, $y_0$, defines the relativistic shifts to radial
frequency and to the period of oscillation. \label{Urlt}}
\end{figure}

We denote $C_r=\tilde{E}_r/m$ the scaled radial energy. We write it in the form
\begin{equation}\label{Cr:E}
C_r=\tilde{{\cal E}}_r^{(0)}+\frac{1}{c^2}C_r^{(1)},
\end{equation}
where the first term in the right-hand side is just
the non-relativistic radial energy introduced in Section
\ref{non-Relat}. The potential in Hamiltonian (\ref{H_qrel}) depends on four parameters $(\omega_c,\omega_z,{\cal E},\ell)$. To make the analysis as clear and concise as possible we define the dimensionless time $t$ and introduce the dimensionless radial variable $\varrho$:
\begin{equation}\label{rho:rel}
t=\Omega_\epsilon\tau,\quad \varrho=\frac{1}{c}\frac{\omega_z^2}{\sqrt{32}\Omega_\epsilon}\tilde{\rho}.
\end{equation}
In terms of these variables the radial Hamiltonian takes the form
\begin{equation}\label{H:rho}
H_\varrho=\frac12p_\varrho^2+U(\varrho),
\end{equation}
where the radial potential is
\begin{equation}\label{U-rho}
U(\varrho) = \frac12\left(\varrho^2+\frac{\lambda}{54}\frac{1}{\varrho^2}-\varrho^4\right).
\end{equation}
Its shape is determined by the dimensionless parameter
\begin{equation}\label{lbd}
\lambda=\frac{1}{c^4}\frac{27}{32}\left(\Omega_\epsilon\ell\right)^2L_\epsilon^4,
\end{equation}
where the factor
\begin{equation}\label{L-epsl}
L_\varepsilon=\frac{\kappa}{1-{\cal E}\kappa}
\end{equation}
can be developed in series up to the first order in powers $1/c^2$:
\begin{equation}\label{L-epskpp}
L_\varepsilon=L_\kappa\left(1+\frac{1}{c^2}L_\kappa{\cal E}^{(0)}\right).
\end{equation}
The other dimensionless parameter
\begin{equation}\label{eps}
\epsilon_\rho=\frac{3}{2c^2}C_rL_\epsilon^2,
\end{equation}
defines the energy level of the Hamiltonian (\ref{H:rho}): $H_\varrho=\epsilon_\rho/12$. Therefore, the motion in $(xy)$-plane is determined by two constants $(\lambda,\epsilon_\rho)$ which play the role of controlling parameters. Figure \ref{Urlt} illustrates the situation.

Putting $p_\varrho=\dot{\varrho}$ in the conserved quantity (\ref{H:rho}) we obtain the first-order differential equation on the radial variable:
\begin{equation}\label{Er}
\dot{\varrho}^2+\varrho^2+\frac{\lambda}{54}\frac{1}{\varrho^2}-\varrho^4
=\frac16\epsilon_\rho.
\end{equation}
Substituting $\varrho^2=y$ we transform it into the following equation
\begin{equation}\label{diffy}
\frac14\dot{y}^2=y^3-y^2+\frac{\epsilon_\rho}{6}y-\frac{\lambda}{54}.
\end{equation}
Factoring the cubic polynomial we present it in the form
\begin{equation}\label{quas-x}
\frac14\dot{y}^2=(y-y_1)(y-y_2)(y-y_0).
\end{equation}
The real and distinct roots can be expressed in terms of trigonometric functions
\begin{equation}\label{yk}
  y_k=\frac{1}{3}\left(1+2\sqrt{1-\epsilon_\rho/2}\cos\phi_k\right),\quad
  k=0,1,2,
\end{equation}
where
\begin{equation}\label{phi_el}
\phi_k=\phi+\frac{2\pi}{3}k,\quad
\phi=\frac13\arccos\frac{1-(3/4)\epsilon_\rho+\lambda/4}{\left(1-\epsilon_\rho/2\right)^{3/2}}.
\end{equation}
The differential equation (\ref{quas-x}) is solved in Ref.~\cite[Sect.IIIB]{YPM15}.
The orbit is parameterized by the squared elliptic sine:
\begin{equation}\label{soleqmr}
y(t)=y_1+(y_2-y_1)\sn^2(\sqrt{y_0-y_1}t-\phi_r\backslash\alpha),
\end{equation}
where constant $\phi_r$ appears as a phase shift.
The modular angle is defined by the roots of cubic polynomial
equation, see \cite[eq. (71)]{YPM15}:
\begin{equation}\label{k_alph}
\sin^2\alpha = \frac{y_2-y_1}{y_0-y_1}=\frac{\sin\phi}{\sin\left(\phi+\pi/3\right)}.
\end{equation}
Besides the well known Handbook \cite{AbrStg} an introduction to
Jacobian elliptic functions and some of their basic relations are
presented in the paper \cite{Ochs}.

Our next task is to derive the characteristics of radial orbit (\ref{soleqmr}).
We expand the argument of inverse trigonometric
function in eq.(\ref{phi_el}) in powers of small parameters
$\epsilon_\rho$ and $\lambda$ which are given by eqs. (\ref{lbd}) and (\ref{eps}).
With the precision sufficient for our purposes we obtain
\begin{equation}\label{phi_el2}
\phi^2=\frac{1}{18}\left[\frac38\epsilon_\rho^2-\lambda+
\frac{\epsilon_\rho}{4}\left(\frac54\epsilon_\rho^2-3\lambda\right)\right].
\end{equation}
Substituting eqs. (\ref{lbd}), (\ref{eps}), and taking into account eqs. (\ref{elll}), (\ref{calE}), and (\ref{Cr:E}) we arrive at
\begin{equation}\label{phi_el1}
\phi=\frac{\sqrt{3}}{8c^2}L_\epsilon^2\tilde{{\cal E}}_r^{(0)}A',
\end{equation}
where
\begin{eqnarray}
A'&=&\left\{A^2
+\frac{2}{c^2}\left[\frac{C_r^{(1)}}{\tilde{{\cal E}}_r^{(0)}}-(1-A^2)\left(\frac{\ell^{(1)}}{\ell^{(0)}}-\frac12L_\kappa{\cal E}^{(0)}\right)\right.\right.\nonumber\\
&+&\left.\left.\frac{1}{16}L_\kappa^2\tilde{{\cal E}}_r^{(0)}\left(1+9A^2\right)\right]
\right\}^{1/2},\nonumber
\end{eqnarray}
and constant $A$ is given by eq.~(\ref{A0}).

Using eqs. (\ref{rho:rel}) we pass to the realistic squared radius
\begin{equation}\label{r2-rho2}
\rho^2=c^2\frac{32\Omega_\epsilon^2}{\omega_z^4}\varrho^2,
\end{equation}
and rewrite the solution (\ref{soleqmr}) as follows:
\begin{equation}\label{sol-qmr}
\rho^2(\tau)=Y_1+(Y_2-Y_1)\sn^2(\Omega_r\tau-\phi_r\backslash\alpha).
\end{equation}
Capital letter $Y_k$ supplemented with subscript index denotes the
turning points obtained from corresponding parameter $y_k$ in
eq.~(\ref{soleqmr}) by the rule (\ref{r2-rho2}). To derive them we
expand the functions (\ref{yk}) and ignore all the terms of higher
order than $c^{-2}$. The calculations are trivial but cumbersome and
we do not bother with details. With the precision sufficient for our
purposes the parameters are
\begin{eqnarray}
Y_1&=&\frac{\tilde{{\cal E}}_r^{(0)}}{\Omega_\varepsilon^2}\Biggl\{1-A'
+\frac{1}{c^2}\left[\frac{C_r^{(1)}}{\tilde{{\cal E}}_r^{(0)}}+\frac{1}{16}L_\kappa^2\tilde{{\cal E}}_r^{(0)}\left(3+6A
+A^2\right)\right]\Biggr\},\label{Y1}
\end{eqnarray}
\begin{eqnarray}
Y_2&=&\frac{\tilde{{\cal E}}_r^{(0)}}{\Omega_\varepsilon^2}\Biggl\{1+A'
+\frac{1}{c^2}\left[\frac{C_r^{(1)}}{\tilde{{\cal E}}_r^{(0)}}+\frac{1}{16}L_\kappa^2\tilde{{\cal E}}_r^{(0)}\left(3-6A
+A^2\right)\right]\Biggr\}.\label{Y2}
\end{eqnarray}
In the non-relativistic approximation $Y_1=a^2/(m\Omega)$ and
$Y_2=b^2/(m\Omega)$ where $a$ and $b$ are semi-axes of the ellipse
pictured in Fig.~\ref{Ellipse}. The zeroth root, $Y_0$, is of order $c^2$ so that the ratios
\begin{equation}\label{Y12:Yo}
\frac{Y_1}{Y_0}=\frac{1}{8c^2}L_\kappa^2{\tilde{{\cal E}}_r^{(0)}}(1-A),\quad
\frac{Y_2}{Y_0}=\frac{1}{8c^2}L_\kappa^2{\tilde{{\cal E}}_r^{(0)}}(1+A).
\end{equation}
The modular angle in eq. (\ref{sol-qmr}) is
\begin{equation}\label{k-alph}
\sin^2\alpha=\frac{\phi}{\sqrt{3}/2}=\frac{1}{4c^2}L_\kappa^2\tilde{{\cal
E}}^{(0)}_rA\,.
\end{equation}
The amplitude of radial oscillation is as follows:
\begin{equation}
Y_2-Y_1=\frac{2\tilde{{\cal E}}_r^{(0)}}{\Omega_\varepsilon^2}
\left(A'-\frac{3}{8c^2}L_\kappa^2\tilde{{\cal E}}_r^{(0)}A\right).\label{Y2Y1}
\end{equation}
And, finally, the radial frequency modified by the special relativity is
\begin{eqnarray}
\Omega_r&=&\sqrt{y_0-y_1}\Omega_\epsilon\label{Omega-r}\\
&=&\Omega\left[1-\frac{1}{2c^2}L_\kappa\left({\cal E}^{(0)}+\frac18L_\kappa\tilde{{\cal E}}_r^{(0)}\left(3-A\right)\right)\right].\nonumber
\end{eqnarray}
In the limit $c\to\infty$ the angle $\alpha$ is equal to zero and the elliptic sine
in eq. (\ref{sol-qmr}) degenerates to trigonometric sine. We obtain the
non-relativistic radial orbit (\ref{r2nr}) divided on the constant $m\Omega$.

The time that the squared radius (\ref{sol-qmr}) needs for a complete cycle
is $T_r=4K(\sin\alpha)/\Omega_r$ where $K(\sin\alpha)$ is the complete elliptic integral
of the first kind \cite[Eq.~(17.3.1)]{AbrStg}:
\begin{eqnarray}
K(\sin\alpha)&=&\int\limits_0^{\pi/2}\frac{{\rm d}\vartheta}{\sqrt{1-\sin^2\alpha\sin^2\vartheta}}\nonumber\\
&\approx&\int\limits_0^{\pi/2}{{\rm d}\vartheta}\left(1+\frac12\sin^2\alpha\sin^2\vartheta\right)\nonumber\\
&=&\frac{\pi}{2}\left(1+\frac{1}{16c^2}L_\kappa^2\tilde{{\cal
E}}^{(0)}_rA\right).
\end{eqnarray}
The real periodicity of the elliptic sine is $4K(\sin\alpha)>2\pi$.

\subsubsection{The in-plane motion}\label{inplane}

In terms of relativistic variables (\ref{rho:rel}) the polar equation (\ref{polar-Nr}) looks as follows:
\begin{equation}\label{polar-Rl}
\frac{{\rm d}\tilde{\varphi}}{{\rm d}t}=\sqrt{\frac{\lambda}{54}}\frac{1}{\varrho^2(t)}.
\end{equation}
We substitute the solution (\ref{soleqmr}) for the squared radius $\varrho^2(t)$ and integrate according
to the definition of the elliptic integral $\Pi\left(\nu;u\backslash\alpha\right)$ of the third kind \cite[Eq.~(17.2.16)]{AbrStg}:
\begin{equation}\label{El-3}
\tilde{\varphi}(u)=\varphi_0+\sqrt{\frac{\lambda}{54(y_0-y_1)}}\frac{1}{y_1}\Pi\left(-n;u\backslash\alpha\right).
\end{equation}
Symbol $u$ denotes the argument $\sqrt{y_0-y_1}t-\phi_r$ of the
elliptic sine in eq.~(\ref{soleqmr}). The absolute value of negative
characteristic $-n$ is the following ratio
\begin{equation}\label{n-Pi}
n=\frac{y_2-y_1}{y_1},
\end{equation}
while the modular angle $\alpha$ is given in eq.~(\ref{k_alph}).

To visualize the in-plane orbit we expand the function $\tilde{\varphi}(u)$ in powers
$v^2/c^2$ and ignore all the terms of higher order than $c^{-2}$. To do it we change the
variables $\sn(u\backslash\alpha)=\sin\vartheta$ in the indefinite integral
defining $\Pi\left(-n;u\backslash\alpha\right)$ and then expand the expression under integral sign:
\begin{eqnarray}
\int\frac{{\rm d}u}{1+n\,\sn^2(u\backslash\alpha)}&=&
\int\frac{{\rm d}\vartheta}{\left(1+n\sin^2\vartheta\right)\sqrt{1-\sin^2\alpha\sin^2\vartheta}}\nonumber\\
&\approx&\int\frac{{\rm d}\vartheta\left(1+\frac12\sin^2\alpha\sin^2\vartheta\right)}{1+n\sin^2\vartheta}\nonumber\\
&=&\frac{1}{\sqrt{1+n}}\left(1-\frac{\sin^2\alpha}{2n}\right)\arctan\left(\sqrt{1+n}\tan\vartheta\right)\nonumber\\
&+&\frac{\sin^2\alpha}{2n}\vartheta.\label{ElintApp}
\end{eqnarray}
Using the equality $y_0y_1y_2=\lambda/54$, eqs.~(\ref{k_alph}) and (\ref{n-Pi}) we calculate the coefficients in the right-hand side of eq.~(\ref{El-3}) where the elliptic integral is approximated by
the right hand side of eq.~(\ref{ElintApp}):
\begin{eqnarray}
\tilde{\varphi}(u)&=&\varphi_0+\sqrt{\frac{y_0}{y_0-y_1}}\left(1-\frac12\frac{y_1}{y_0-y_1}\right)
\arctan\left(\sqrt{\frac{y_2}{y_1}}\tan\vartheta\right)\nonumber\\
&+&\frac12\frac{\sqrt{y_0y_1y_2}}{(y_0-y_1)^{3/2}}\vartheta.
\end{eqnarray}
Using eqs. (\ref{Y12:Yo}) we prove that the factor before the inverse trigonometric function is approximately equal to unit. We denote $\upsilon$ the factor before $\vartheta$:
\begin{equation}\label{upsil}
\upsilon=\frac{1}{16c^2}L_\kappa^2\left(\Omega\ell^{(0)}\right).
\end{equation}
We substitute $\sin\vartheta=\sn(u\backslash\alpha)$, $\cos\vartheta=\cn(u\backslash\alpha)$, and $\vartheta=\am(u\backslash\alpha)$ and take into account that the amplitude $\am(u\backslash\alpha)\approx u$ with the precision sufficient for our purposes. To visualize the orbit we apply the algorithm presented in Section~\ref{non-Relat} (see eqs.~(\ref{varphi-nr})-(\ref{xy-nr})).
We finally obtain
\begin{eqnarray}\label{xy-Rl}
\left[\begin{array}{c}
\tilde{\xi}(u)\\
\tilde{\chi}(u)
\end{array}
\right]&=&\left[
\begin{array}[c]{cc}
\cos\left(\varphi_0+\upsilon u\right)&-\sin\left(\varphi_0+\upsilon u\right)\\
\sin\left(\varphi_0+\upsilon u\right)&\cos\left(\varphi_0+\upsilon u\right)
\end{array}
\right]\left[\begin{array}{c}
\sqrt{y_1}\cn(u\backslash\alpha)\\
\sqrt{y_2}\sn(u\backslash\alpha)
\end{array}\right],
\end{eqnarray}
where $u=\sqrt{y_0-y_1}t-\phi_r$. In the limit $c\to\infty$ the Jacobian
elliptic functions degenerate to trigonometric functions and we
obtain the orbit (\ref{xy-nr}).

\begin{figure}[ht]
\begin{center}
\includegraphics*[scale=0.75,angle=0,trim=1 5 1 3]{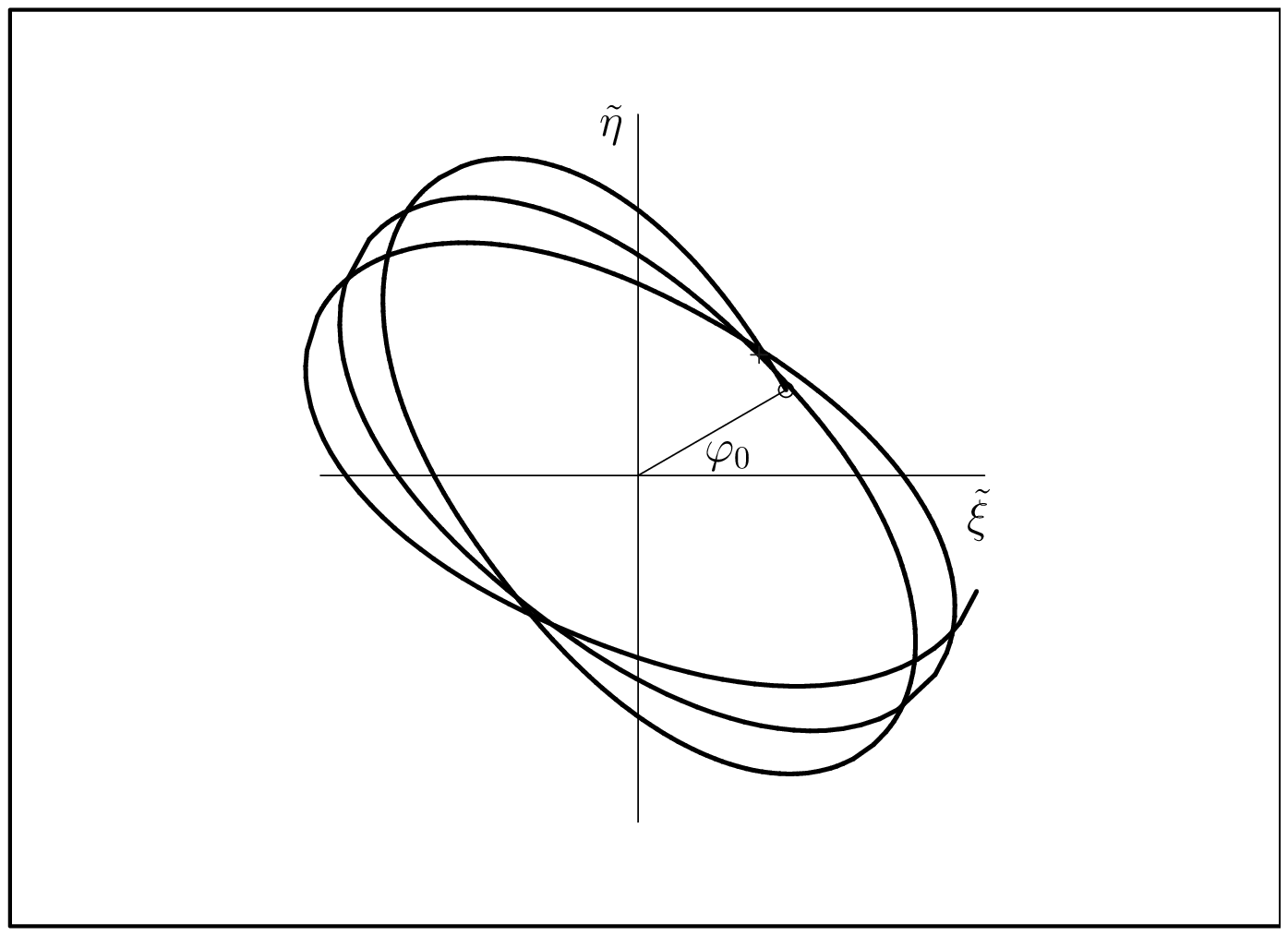}
\end{center}
\caption{Relativistic corrections yield precession of particle's in-plane orbit.
The rate (\ref{precess}) of this precession is proportional to the infinitesimal
parameter (\ref{upsil}) which defines the shift of the argument of the rotational
matrix involved in eq.~(\ref{xy-Rl}). The starting point (circle) does not coincide with the cross point after period $T=4K(\sin\alpha)$ of oscillation.
\label{Ellipse-Rl}}
\end{figure}

What is the trajectory given by eq.~(\ref{xy-Rl})? In the non-relativistic
problem a charge follows the same perfect ellipse constantly.
If we take into account the relativistic corrections to in-plane motion
the ellipse gradually rotates. Figure \ref{Ellipse-Rl} illustrates the
situation. The reason is the time-dependent term $\upsilon u$ which arises
in the arguments of of trigonometric functions which constitute the rotational
matrix in eq.~(\ref{xy-Rl}). During the period $T=4K(\sin\alpha)$ that
the $\cn$ and $\sn$ functions
need for a complete cycle the angle at which ellipse's axes are inclined to
coordinate axes changes on $\upsilon\cdot 4K(\sin\alpha)$. With the precision
$1/c^2$ the shift of this tilt angle is:
\begin{equation}\label{precess}
\triangle\varphi=\frac{2\pi}{16c^2}L_\kappa^2\left(\Omega\ell^{(0)}\right).
\end{equation}
To establish how the in-plane orbit looks in the laboratory frame we
pass to the $\tilde{\rho}$ and $\tau$ inverting eqs.~(\ref{rho:rel})
and perform the coordinate transformation (\ref{xy_nr}). We obtain a
rosette shape curve similar to that pictured in Fig.~\ref{Rosette}.

\subsection{Axial motion}\label{Axial}

We denote $C_z=\tilde{E}_z/m$ the scaled axial energy which is equal to the total energy (\ref{calEtld}) minus the radial energy (\ref{Cr:E}):
\begin{equation}\label{Cz:E}
C_z=\tilde{{\cal E}}_z^{(0)}+\frac{1}{c^2}C_z^{(1)}.
\end{equation}
The first term in the right-hand side is just the non-relativistic axial energy introduced in Section \ref{non-Relat}. We introduce the dimensionless variables
\begin{equation}\label{ax:rel}
t={\cal E}\omega_z\tau,\quad \zeta=\frac{1}{c}\frac{\omega_z}{\sqrt{6{\cal E}}}z,
\end{equation}
where ${\cal E}$ is the total energy (\ref{calE}) in the laboratory reference frame.

In terms of these variables the axial Hamiltonian takes the form
\begin{equation}\label{H:ksi}
H_\zeta=\frac12p_\zeta^2+V(\zeta),
\end{equation}
where the axial potential is
\begin{equation}\label{V-ksi}
V(\zeta) = \frac12\left(\zeta^2-\zeta^4\right).
\end{equation}
The energy level of this Hamiltonian $H_\zeta=\epsilon_z/8$ is defined by the dimensionless controlling parameter
\begin{equation}\label{eps-z}
\epsilon_z = \frac{1}{c^2}\frac{4C_z}{3{\cal E}^2}.
\end{equation}
Putting $p_\zeta=\dot{\zeta}$ in the conserved quantity (\ref{H:ksi}) we obtain the first-order differential equation on the axial variable:
\begin{equation}\label{Ez}
\dot{\zeta}^2+\zeta^2-\zeta^4
=\frac14\epsilon_z.
\end{equation}
We write the axial equation (\ref{Ez}) in the standard form
\begin{equation}\label{eq1-zJ}
\dot{\zeta}^2=\frac14\epsilon_z\left(1-\frac{\zeta^2}{a_z^2}\right)\left(1-\frac{\zeta^2}{b_z^2}\right),
\end{equation}
where
\begin{equation}\label{ab}
a_z^2=\frac12\left(1-\sqrt{1-\epsilon_z}\right),\quad b_z^2=\frac12\left(1+\sqrt{1-\epsilon_z}\right).
\end{equation}
Substituting $\zeta=a_z\sin\vartheta$ we clearly recognize the equation on elliptic integral of the first kind:
\begin{equation}\label{Eqn:z}
\dot{\vartheta}^2=b_z^2\left(1-k^2\sin^2\vartheta\right).
\end{equation}
The differential equation can be inverted and put in integral form \cite[Eq.~17.2.6]{AbrStg}.
The elliptic modulus $k=a_z/b_z$ is a real number $0\leq k<1$:
\begin{equation}\label{k:z}
k=\sqrt{\frac{1-\sqrt{1-\epsilon_z}}{1+\sqrt{1-\epsilon_z}}}.
\end{equation}
The non-linear equation is not limited to describe the axial motion but also the motion of a gravity pendulum \cite[\S 3.3]{BB} (see also the references therein). The dynamics is analyzed in details, including driven systems and chaos. The equation (\ref{Eqn:z}) describes the pendulum which does not possess sufficient energy for a complete cycle \cite[Sect.5]{Ochs}.

\begin{figure}[ht]
\begin{center}
\includegraphics*[scale=0.7,angle=0,trim=1 3 2 1]{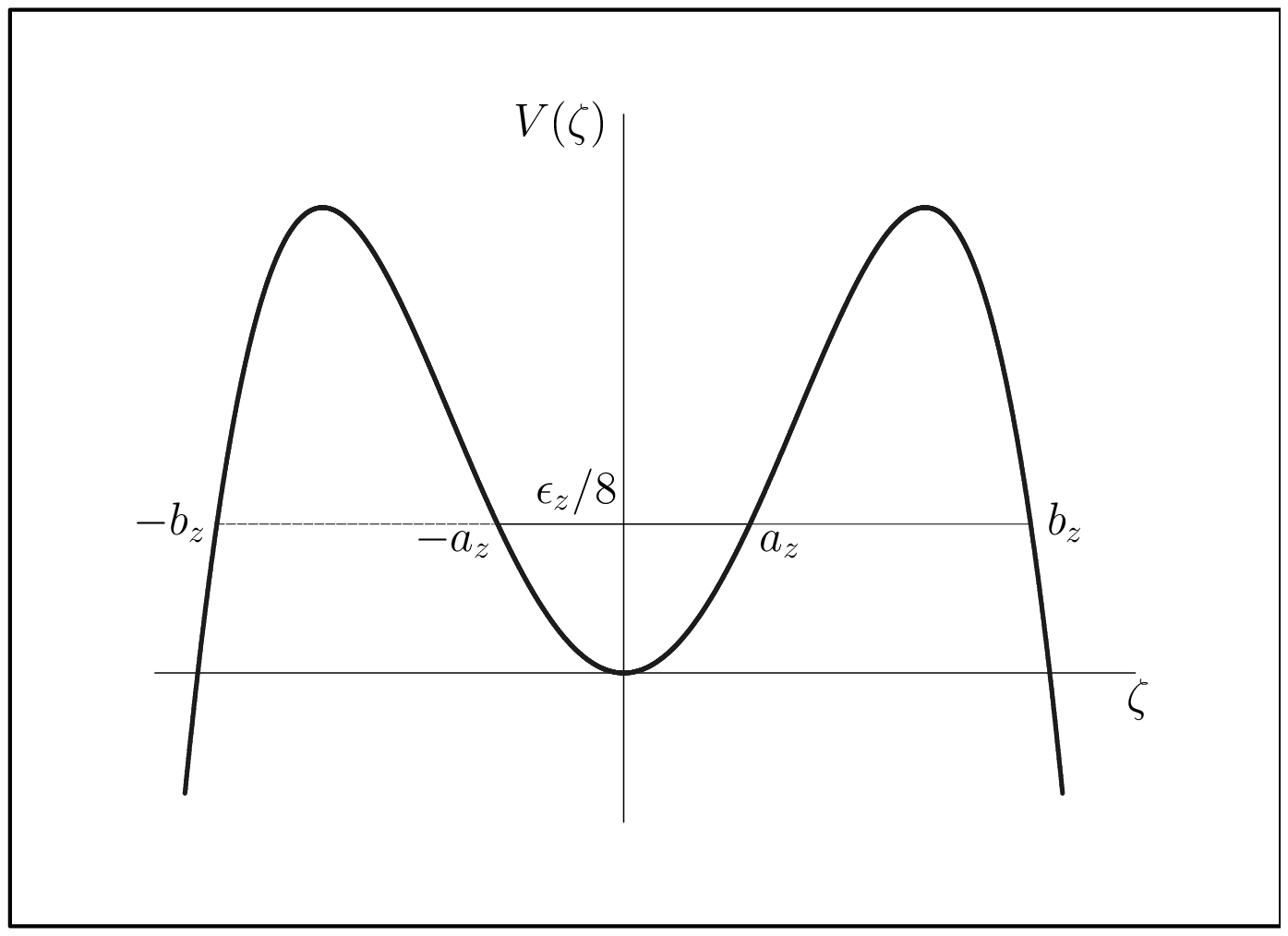}
\end{center}
\caption{Graph of the relativistic axial potential (\ref{V-ksi}). For a fixed value $\epsilon_z\ll 1$ the
axial variable oscillates between the roots $-a_z$ and $a_z$ of the
algebraic equation $V(\zeta)=\epsilon_z/8$. The absolute value of the roots $\pm b_z$ defines the
frequency and the elliptic modulus. The low limit $\epsilon_z^{\rm min}=4(\omega_-l^{(0)})/3c^2$ corresponds to the minimal value (\ref{Ez-min}) of the non-relativistic axial energy. \label{V:Axial}}
\end{figure}

The solution to eq. (\ref{eq1-zJ})
\begin{equation}\label{axial:z}
\zeta(t)=a_z\sn\left(b_z\tau-\phi_z|k\right)
\end{equation}
describes the oscillation of axial variable near the coordinate origin with constant amplitude $a_z$ and frequency $b_z$. The periodicity $4K(k)>2\pi$. Figure \ref{V:Axial} illustrates the situation.

Using eqs. (\ref{ax:rel}) we rewrite the solution (\ref{axial:z}) in terms of the realistic variables:
\begin{equation}\label{axial-z}
z(\tau)=A_z\sn\left(\Omega_z\tau-\phi_z|k\right).
\end{equation}
We expand the characteristics of orbit in series up to the first order in powers $1/c^2$.
With the precision sufficient for our purposes the amplitude
\begin{equation}
A_z=\frac{\sqrt{2\tilde{{\cal E}}_z^{(0)}}}{\omega_z}
\Biggl[1+\frac{1}{2c^2}\Biggl(\frac13\tilde{{\cal E}}_z^{(0)}
-{\cal E}^{(0)}+\frac{C_z^{(1)}}{\tilde{{\cal E}}_z^{(0)}}\Biggl)\Biggr],\label{A-z}
\end{equation}
and modified frequency
\begin{equation}\label{Omega-z}
\Omega_z=\omega_z\Biggl[1+\frac{1}{2c^2}\left({\cal
E}^{(0)}-\frac13\tilde{{\cal
E}}_z^{(0)}\right)\Biggr].
\end{equation}
Since the elliptic modulus is of order $c^{-2}$
\begin{equation}\label{kk-z}
k^2=\frac{1}{3c^2}\tilde{{\cal E}}_z^{(0)},
\end{equation}
the period of
oscillation $T_z=4K(k)/\Omega_z$ can be approximated as
\begin{equation}\label{Period-z}
T_z=\frac{2\pi}{\Omega_z}\left(1+\frac{1}{12c^2}\tilde{{\cal E}}_z^{(0)}\right).
\end{equation}

\subsubsection{Minimum of the radial potential}\label{min-x2}

Let us consider the specific situation when the radial variable
is equal to the point at which the potential (\ref{U-rho}) takes minimal value. This orbit is usual for electron
because the emission of synchrotron radiation suppresses the
fast-oscillating radial mode \cite{GD85}.
The radial coordinates of critical points satisfy the algebraic equation ${\rm d}U(\varrho)/{\rm d}\varrho=0$.
Substituting $\rho^2=y$ we transform it into the cubic polynomial equation
\begin{equation}\label{cube:r}
y^3-\frac12y^2+\frac{\lambda}{108}=0.
\end{equation}
We express the roots in terms of trigonometric functions:
\begin{equation}\label{root:r}
x_k=\frac16\left(1+2\cos\psi_k\right),\quad k=0,1,2,
\end{equation}
where
\begin{equation}\label{root:psi}
\psi_k=\psi+\frac{2\pi}{3}k,\quad \psi=\frac13\arccos\left(1-\lambda\right).
\end{equation}
The root $x_2$ is the local minimum of potential (\ref{U-rho}) while $x_0$ is the local maximum (see Fig. \ref{Urlt}). The third root $x_1$ is negative, so it does not correspond to any real solution for $\rho$.

To obtain the realistic coordinate of minimum we substitute $x_2$ for $\varrho^2$ in eq. (\ref{r2-rho2}) and develop the function in series up to the first order in powers $1/c^2$:
\begin{equation}\label{r2min}
X_2=\frac{\ell^{(0)}}{\Omega}\left\{1+\frac{1}{c^2}
\frac{\ell^{(1)}}{\ell^{(0)}}+\frac{L_\kappa}{2c^2}
\left[\tilde{{\cal E}}_z^{(0)}-(\ell^{(0)}\omega_-)\frac{\omega_+-2\omega_-}{2(\omega_+-\omega_-)}\right]\right\}.
\end{equation}
We take into account the relation for minimal energy
${\cal E}_{\rm min}^{(0)}=\tilde{{\cal E}}_z^{(0)}-\ell^{(0)}\omega_-$
derived in Section \ref{non-Relat} (see eq. (\ref{Ez-min})).

At point of minimum the radial velocity $\dot{\rho}=0$ and the radial energy $C_r$ takes the minimal value
\begin{eqnarray}
C_r^{\rm min}&=&\frac{2c^2}{3L_\epsilon^2}\epsilon_\rho^{\rm min}\nonumber\\
&=&\frac{8c^2}{L_\epsilon^2}x_2\left(1-\frac{3}{2}x_2\right).
\end{eqnarray}
With the precision sufficient for our purposes
\begin{eqnarray}\label{Er_min}
C_r^{\rm min}&=&(\ell^{(0)}\Omega)\left\{1+\frac{1}{c^2}
\frac{\ell^{(1)}}{\ell^{(0)}}\right.\nonumber\\&-&\left.\frac{1}{2c^2}(\ell^{(0)}\Omega)L_\kappa
\left[\tilde{{\cal E}}_z^{(0)}-
(\ell^{(0)}\omega_-)\frac{3\omega_+-4\omega_-}{4(\omega_+-\omega_-)}\right]\right\}.
\end{eqnarray}
At this energy level the turning points (\ref{Y1}) and (\ref{Y2}) coincide
and equal to $X_2$. Particle's trajectory is the combination of circular
magnetron motion with radius $r_{\rm min}=\sqrt{X_2}$ and oscillation along the $z$-axis.
Characteristics of axial orbit (\ref{axial-z}) are given by
eqs.~(\ref{A-z})-(\ref{Period-z}) where the minimal radial energy is inserted.

\subsection{Laboratory time}\label{tau-t}

A charged particle oscillates in according to its proper time, while
a researcher measures the laboratory time. To evaluate characteristics
of particle's motion properly, the expression which relates these evolution
parameters is necessary. To derive the one-to-one correspondence between
$\tau$ and $t$ we solve the first order differential equation (\ref{u0vphi:E}).
With the precision sufficient for our purposes we substitute the
non-relativistic approximations of radial and axial orbits for
$r^2$ and $z^2$, respectively:
\begin{eqnarray}
\frac{{\rm d}x^0}{{\rm d}\tau}&=&1+\frac{1}{c^2}{\cal
E}^{(0)}+\frac{\omega_z^2}{2c^2}\left\{\frac{\tilde{{\cal
E}}_r^{(0)}}{2\Omega^2}\Bigl[1-A\cos\left(2\Omega\tau-2\phi_r\right)\Bigr]\right.\nonumber\\
&-&\left.\frac{\tilde{{\cal
E}}_z^{(0)}}{\omega_z^2}\Bigl[1-\cos\left(2\omega_z\tau-2\phi_z\right)\Bigr]\right\}.\nonumber
\end{eqnarray}
The solution is the combination of linear term and trigonometric functions:
\begin{eqnarray}
\frac{\tau}{c}&=&t\left\{1-\frac{1}{c^2}\left[\frac12\tilde{{\cal E}}_z^{(0)}+\tilde{{\cal E}}_r^{(0)}\left(1+\frac12L_\kappa\right)-\frac12\omega_c\ell^{(0)}\right]\right\}\nonumber\\
&+&\frac{1}{4c^2}\left\{\frac{\tilde{{\cal
E}}_r^{(0)}A}{\Omega}L_\kappa\Bigl[\sin(2\Omega t-2\phi_r)+\sin
2\phi_r\Bigr]\right.\nonumber\\
&-&\left.\frac{\tilde{{\cal
E}}_z^{(0)}}{\omega_z}\Bigl[\sin(2\omega_zt-2\phi_z)+\sin
2\phi_z\Bigr]\right\}.\label{tau_t}
\end{eqnarray}
The proper time flows slower than the laboratory time. The scaled proper time $\tau/c$ plays the role of the evolution parameter in dynamics produced by the Hamiltonian (\ref{H_qrel}) (the factor $1/c$ is omitted in subsequent formulae).

\section{Discussion and conclusions}\label{Concl}
\setcounter{equation}{0}

Adding the precisely tuned octupolar potential we cancel the term $\rho^2z^2$ in the relativistic corrections to the perfect quadru\-po\-le potential and improve the relativistic equations of motion of a single ion in an ideal Penning trap. For the uncoupled oscillating modes, radial and axial, we derive the first-order non-linear differential equations which are analogous to that governing the motion of the simple gravity pendulum
\cite{BB,Ochs}. The solutions are expressed in terms of Jacobian elliptic functions. We restrict our consideration to the first order in $v^2/c^2$ where $v$ is the velocity of charge and $c$ is speed of light. As the parameter $m=\sin^2\alpha=k^2$ is so small that we may neglect $m^2$ and higher powers, the Jacobian elliptic functions can
be approximated by trigonometric functions \cite[\S 16.13]{AbrStg}:
\begin{eqnarray}
\sn(u|m)&\approx&\sin u-\frac14m\left(u-\sin u\cos u\right)\cos u,\label{sn:trig}\\
\cn(u|m)&\approx&\cos u+\frac14m\left(u-\sin u\cos u\right)\sin u.\label{cn:trig}
\end{eqnarray}
On the basis of these formulae we can rewrite the radial oscillating mode
(\ref{sol-qmr}) and axial oscillating mode (\ref{axial-z}) which are
compatible with expressions obtained previously in Ref.~\cite{Kett14}. To get
coincidence between two approaches we should replace the proper time by
the laboratory time which is used in Ref.~\cite{Kett14}. The expression
(\ref{tau_t}) which relates the evolution parameters is also the combination of
linear term and trigonometric functions.

In our opinion, the conceptual framework of description of relativistic
motion of a charge in a Penning trap involves both the proper time parametrization
and the elliptic Jacobian functions. The particle's own clock shows an earlier time
than the laboratory time. But a charged particle follows the periodic orbit according to
the particle's proper time. To evaluate eigenfrequencies properly we should use the
particle's proper time as the evolution parameter. Secondly,
the relativistic corrections make the equations of nonlinear even in the first order
in small parameter $v^2/c^2$. Similarly, the linearized differential equation
describes periodic oscillation of a low energy simple gravity pendulum.
The period is independent of amplitude, i.e. on the total energy of oscillator.
If the energy of pendulum increases we should solve the nonlinear equation to describe
the oscillation properly because the period increases gradually with amplitude \cite[Fig.~3.17]{BB}.

Usage of the particle's proper time instead of the laboratory time and Jacobian elliptic functions instead of ordinary trigonometric functions will increase the accuracy of measurements of the relativistic shifts of eigenfrequencies. As the real period of oscillation of a charged particle exceeds $2\pi$, the systematic error accumulates with time whenever we parameterize the periodic process by trigonometric functions. Indeed, the terms $u\cos u$ and $u\sin u$ involved in eqs. (\ref{sn:trig}) and (\ref{cn:trig}) arise in a perturbation-series solution of the Duffing equation which illustrates secular (i.e., long-term) influence of interplanetary gravitational perturbations on planetary orbits.

The anharmonic axial resonance \cite[Sect. III.D]{BB} can reveal the impact of the new calculations for measurements. The electrodes are designed for producing the electrostatic potential (\ref{Phi24}) which cancels the cross term $\rho^{2}z^{2}$ in the relativistic effective potential (\ref{V}) while the terms $z^4$ and $\rho^4$ survive. The axial and radial motions become uncoupled and the noise vanishes even if the driving force is relatively small.

\section*{Acknowledgement}

The author gratefully acknowledges stimulating comments of unknown reviewers who propose
to state the electrostatic potential that emphasizes the effects of special relativity.
This research has been supported by Grant No 0116U005055 of the
State Fund For Fundamental Research of Ukraine.

\end{document}